# The answer is blowing in the wind

Yousaf M. Butt

**A source of astoundingly energetic γ-rays associated with a star cluster might provide a clue to a century-old question: where do the cosmic rays that constantly bombard Earth come from?**

Massive stars have extreme lifestyles. They are born in clusters of up to several thousand members, blow fierce charged-particle winds during their short lives, and die — more or less together — in powerful supernova explosions. Now comes word from the High Energy Stereoscopic System (HESS) collaboration, to be published in *Astronomy and Astrophysics*[1], that γ-rays of very high energy have been spotted coming from the powerful young stellar association Westerlund 2 located in the southern sky[1] (Fig 1). This emission is of a higher energy than ever seen before from a group of stars, and pushes the limits of our understanding of the processes behind it.

Stars typically emit light around the visible part of the spectrum, where photons have an energy of a few electronvolts (eV). The γ-rays that HESS detected have energies in the range of tera-electronvolts (TeV), or $10^{12}$ eV. Previously, TeV γ-rays have been seen emanating from only a handful of exotic celestial objects. These include energetic pulsars (rapidly spinning and highly magnetized neutron stars just 30 or so kilometres across); the huge interstellar shock waves associated with the remnants of powerful supernovae; binary systems of a neutron star or a black hole coupled with a regular star; jets from distant 'active galaxies'; and the supermassive black hole thought to lurk at the centre of our Galaxy. So the HESS collaboration's discovery[1], dubbed HESS J1023–575, amounts to finding a completely new species of celestial γ-ray source.

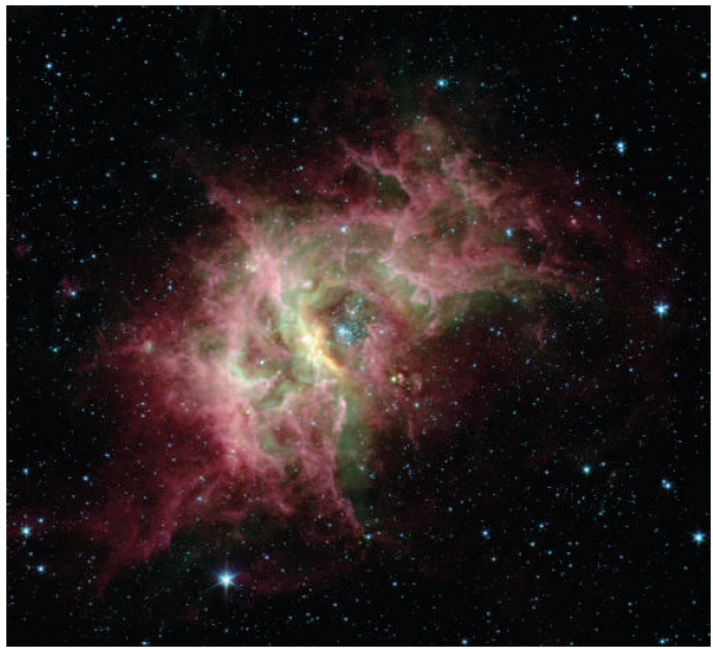

Figure 1 *Turbulent association*. A composite infrared image from NASA's orbiting Spitzer observatory shows the Westerlund 2 stellar association. According to observations from the HESS telescope[1], it seems that turbulent processes at work in the fierce winds of the massive stars in Westerlund 2 create γ-ray photons $10^{12}$ times more energetic than visible light. The same processes might cause the acceleration of the cosmic rays that constantly hit Earth.

In fact, another TeV source discovered recently[2] might also be a member of the same species. Designated TeV J2032+4130, this source is probably related to a subgroup of powerful stars in the Cygnus OB2 stellar association[3], but this identification is not quite as firm as in the case of Westerlund 2.

The most likely model for the origin of these highly energetic γ-rays is that multiple, supersonic winds of charged particles blowing from the dozens of massive stars (for our purposes, stars bigger than 8 solar masses) create violent plasma motions within Westerlund 2. This turbulence can accelerate particles to TeV energies (ref. 4 and references therein), and these particles can then interact with the ambient material and light to produce the detected γ-rays. This type of turbulent particle acceleration process is called the second-order Fermi mechanism, or Fermi-II acceleration for short. First-order Fermi (Fermi-I) acceleration is thought to be at work in the better-formed interstellar shock waves created by isolated supernova explosions.

Could such an isolated supernova remnant be behind the HESS J1023–575 detection? This possibility is rendered unlikely by the presence of a great deal of turbulence caused by the massive stars of the Westerlund 2 association. The evolution of a supernova remnant would be greatly perturbed in such an environment, and it could hardly be considered as 'isolated'.

On the other hand, the possibility that one supernova remnant or more could have added to the turbulence created in Westerlund 2 by the massive stars resident there, and thus provided a further power source for the Fermi-II process most probably operating there, certainly can't be discounted. Although Westerlund 2 is a young stellar association (it is roughly 2 million years old), massive stars 'live fast and die young': they burn quickly and brightly. A supernova might well, therefore, already have occurred in Westerlund 2 — which might also explain a deficit discovered[5] in counts of the most massive members of Westerlund 2. There is no tell-tale sign of any supernova remnant in the area, but we should not necessarily expect there to be[6]: Westerlund 2's medium is so hot and rarefied that any remnant would leave a barely detectable signature.

This brings us to the most important aspect of the HESS findings[1]: their bearing on the origin of cosmic rays. Cosmic rays are very energetic particles — mostly protons, but also heavier nuclei and electrons — that continually rain down on Earth's atmosphere from space. No one knows for sure where in our Galaxy, or beyond, they are accelerated.

It is widely believed that supernova remnants are the main energy source powering the acceleration of cosmic rays. But the acceleration mechanism itself can be quite different, depending on whether the remnant is isolated (where Fermi-I acceleration dominates) or is part of an interacting system of several remnants embedded in a turbulent medium such as a stellar association (mainly Fermi-II acceleration)[4,7]. Because the massive stars that lead to supernovae are born together in stellar clusters, roughly 80% of supernova explosions are expected to take place near others within a relatively short period of time[8,9] — much like a closing fireworks display — creating a large 'superbubble' filled with hot, tenuous, turbulent plasma[10–13].

Cosmic rays, it has been argued[4,7–9], are much more likely to be accelerated in such superbubbles predominantly via the Fermi-II process, rather than — or, possibly, in addition to — in isolated supernova remnants where the Fermi-I mechanism holds sway. The HESS findings[1], which seem to confirm the viability of turbulent, Fermi-II acceleration, whether by massive stars alone or by supernova remnants in concert with massive stars, will be welcome news to the proponents of the superbubble mechanism of cosmic-ray acceleration. But there is considerable work still ahead.

At the theoretical end, it would be useful to develop the details of superbubble models[4], such that concrete predictions of emissions from realistic sources, from radio to TeV energies, can be made. Observationally, we must now aim to go beyond examining only the isolated supernova remnants as possible cosmic-ray acceleration sites. Even if the intriguing TeV-emitting isolated supernova remnants that are already known can be shown to be accelerating cosmic rays, it would not solve the general cosmic-ray acceleration problem for two reasons. First, the well-known and catalogued isolated supernova remnants are a distinct minority of the supernova remnants in the Galaxy. Second, the precise mechanisms of acceleration at work in the isolated supernova remnants and superbubbles are probably different.

Luckily, a suite of sensitive ground- and space-based γ-ray observatories with acronyms such as VERITAS, MAGIC, GLAST and AGILE is now coming online. Together with HESS, these hold the promise of a rich harvest of celestial γ-ray sources. Detecting the large, diffuse and possibly overlapping superbubbles will be very challenging and time-consuming, and perhaps downright impossible with some current instrumentation. But such objects ought to be pursued with these and other lower-frequency observatories if we are to understand how and where cosmic rays are accelerated. ∎

*Yousaf M. Butt is at the National Centre for Physics, Quaid-i-Azam University, Islamabad, Pakistan and at the Harvard-Smithsonian Center for Astrophysics, Cambridge, Massachusetts 02138, USA.*

e-mail: ybutt@cfa.harvard.edu